\documentclass[a4paper,11pt]{article}
\usepackage{pos}
\usepackage{wrapfig}
\graphicspath{{logos/},{fig/}}

\title{ITS3: A truly cylindrical inner tracker for ALICE}
\author*[a]{Jory Sonneveld for the ALICE Collaboration}
\affiliation[a]{Nikhef,\\
  Science Park 105, Amsterdam, the Netherlands}

  \emailAdd{jory.sonneveld@nikhef.nl}

  \abstract{After the successful installation and first operation of the new Inner Tracking System (ITS2), which consists of about 10 m$^2$ of monolithic silicon pixel sensors, ALICE is pioneering the usage of bent, wafer-scale pixel sensors for the ITS3 for Run 4 at the LHC in 2029. Sensors larger than typical reticle sizes can be produced using the technique of stitching. At thicknesses of about 30 $\mu$m, the silicon is flexible enough to be bent to radii of the order of 1 cm. By cooling such sensors with a forced air flow, it becomes possible to construct a detector with minimal material budget. The reduction of the material budget and the improved pointing resolution will allow new measurements, in particular of heavy-flavor decays and electromagnetic probes. Mechanical studies have shown the sensors to be unaffected by bending, and bent sensors have been shown to be fully efficient in test beams. New sensor developments for the ITS3 have shown promising results for fluences even beyond those expected for ITS3.}

\FullConference{% HardProbes2023\\
 26-31 March 2023 \\
 Aschaffenburg, Germany\\}

\begin{document}
\maketitle

The current ALICE inner tracking system (ITS2) 
% as shown on the CERN Courier in Fig. \ref{fig:pixelperfect}, 
is the first with monolithic active pixel sensors (MAPS) at the Large Hadron Collider (LHC) at CERN \cite{cerncourier}. It consists of 3 inner layers at distances from 22 mm to 42 mm from the interaction point (IP) that have a material budget of only 0.36\% $X_0$. The outer tracker consists of 4 layers of MAPS at 194 mm to 395 mm from IP with a material budget of 1.1\% $X_0$. It features ALICE PIxel DEtectors (ALPIDEs) \cite{alpide} with $27\times29$~$\mu$m$^2$ pixels. With 12.5 Gigapixels and 10 square meters active area, it is the largest pixel detector built to date. It has successfully taken data since September 2021.
%\begin{wrapfigure}{1}{0.21\textwidth}
%    \centering
%    \includegraphics[width=\linewidth]{fig/cerncourier.png}
%    \caption{ALICE Inner Tracking System 2 (ITS), outer barrel, as shown on the CERN Courier \cite{cerncourier}.}
%    \label{fig:pixelperfect}
%\end{wrapfigure}
\begin{figure}
    \centering
    \includegraphics[width=0.52\textwidth,  trim=5 10 5 5, clip]{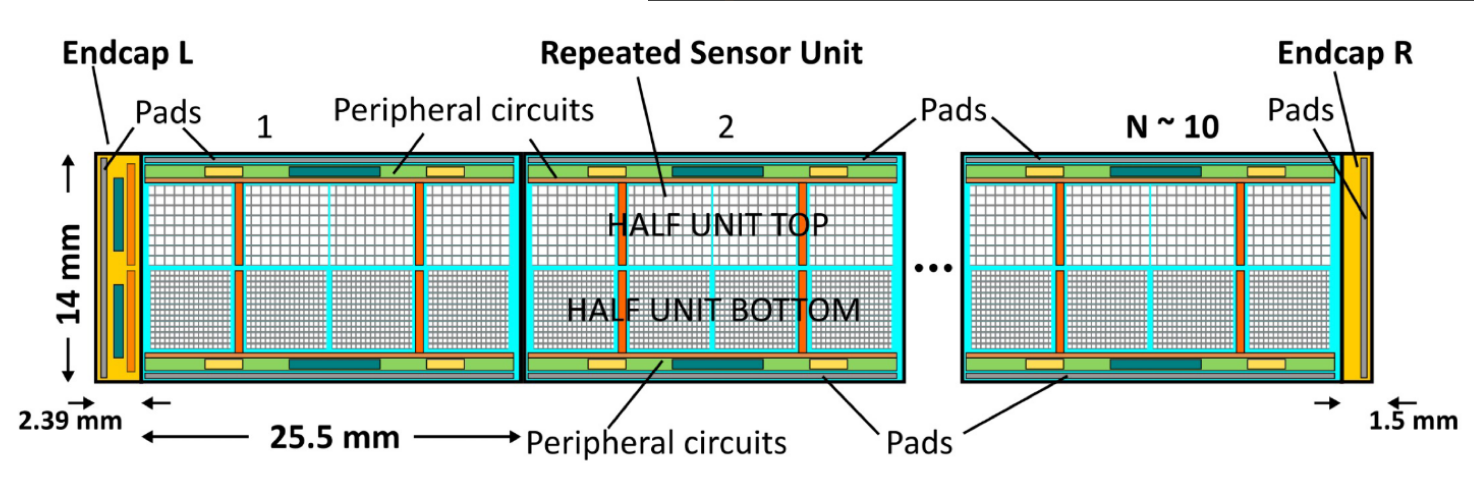}
    \includegraphics[width=0.47\textwidth,  trim=5 10 5 5, clip]{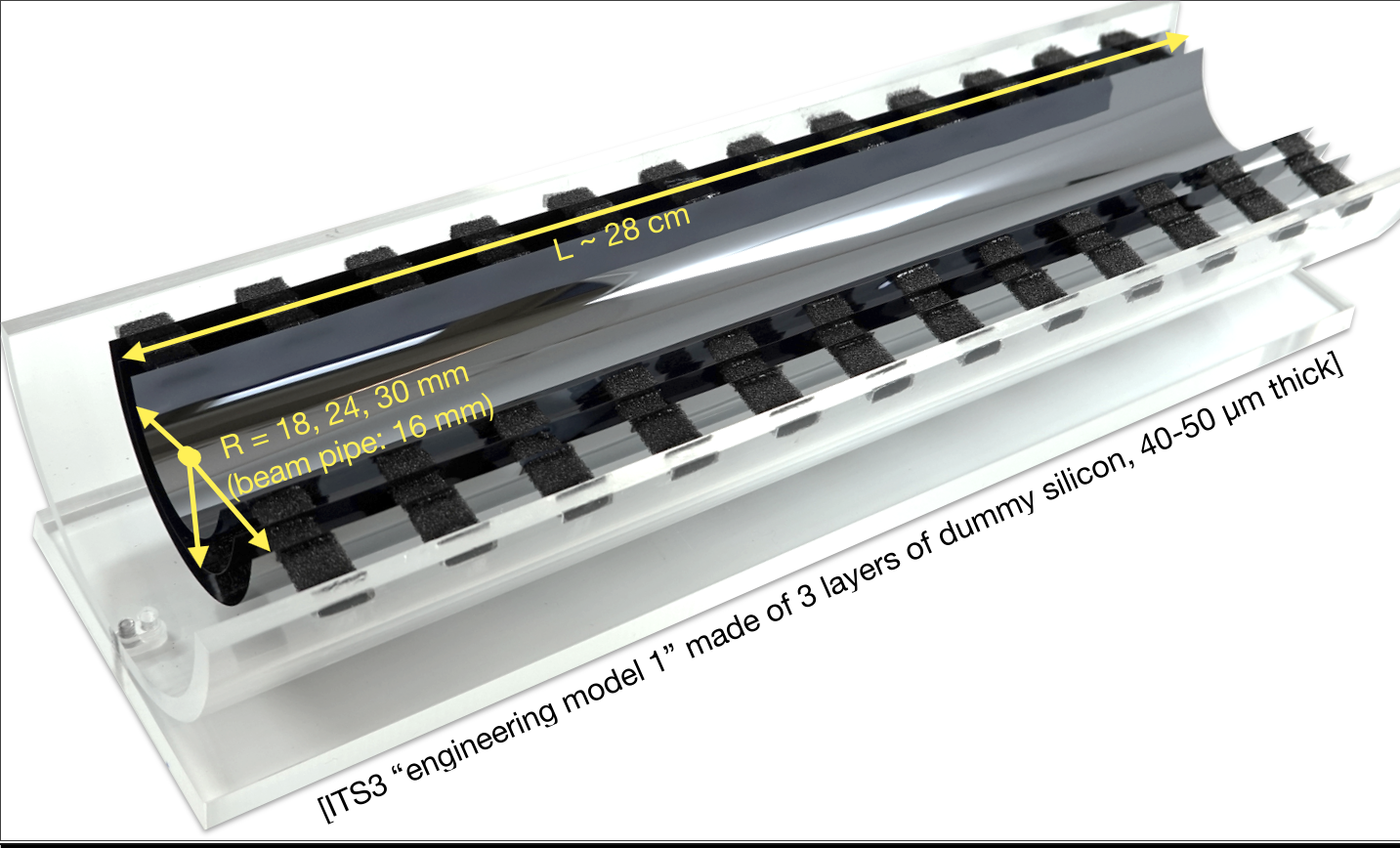}
    \caption{
        %Left: a 25.9 cm pad wafer using stitching (middle lanes) from engineering run 1.
        Left: layout of a monolithic stitched sensor prototype.
Right: ITS3 engineering model 1 made of 3 layers of 40-50 $\mu$m thick dummy silicon. Figure from \cite{magnusdrd7}.}
    \label{fig:er1em1}
\end{figure}

\begin{wrapfigure}{1}{0.21\textwidth}
    \centering
    \includegraphics[width=\linewidth]{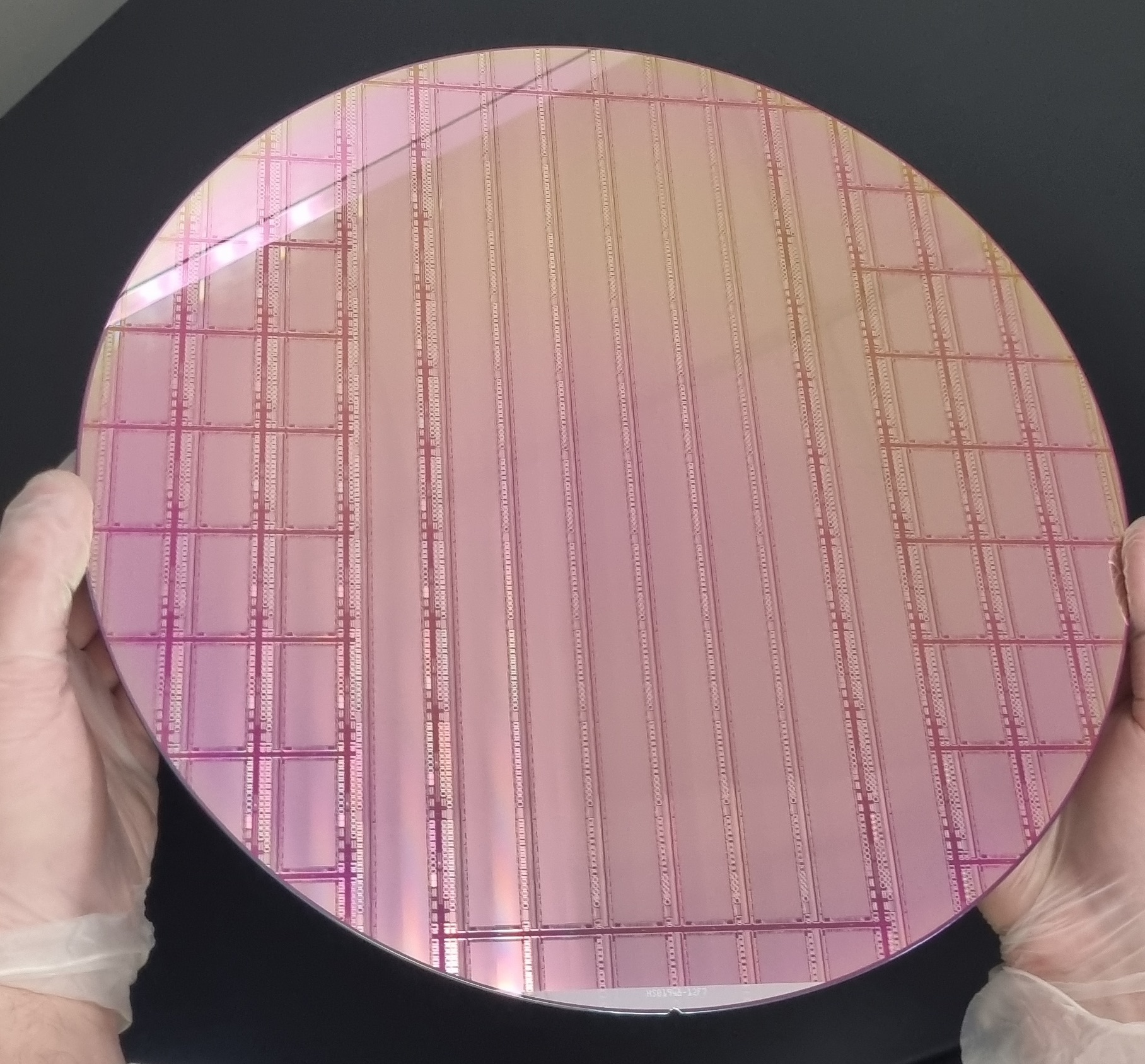}
    \caption{
    A 30 cm pad wafer using stitching (middle lanes) from engineering run 1.}
    \label{fig:stitchedsensor}
\end{wrapfigure}
\section{The upgrade of the ALICE inner tracking system for Run 4}
\begin{figure}
    \centering
\includegraphics[width=0.34\textwidth, trim=677 10 5 5, clip]{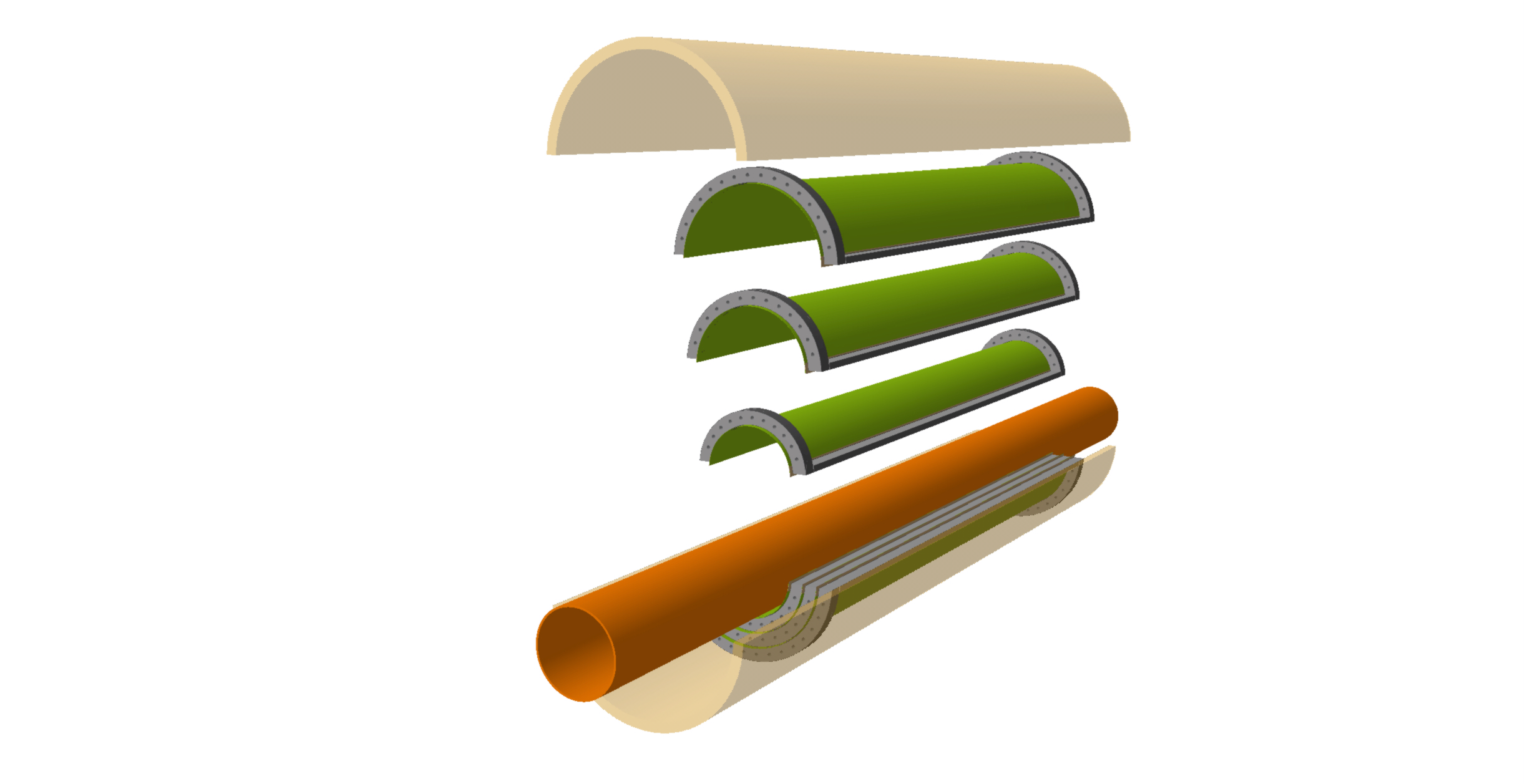}
    \includegraphics[width=0.474\textwidth]{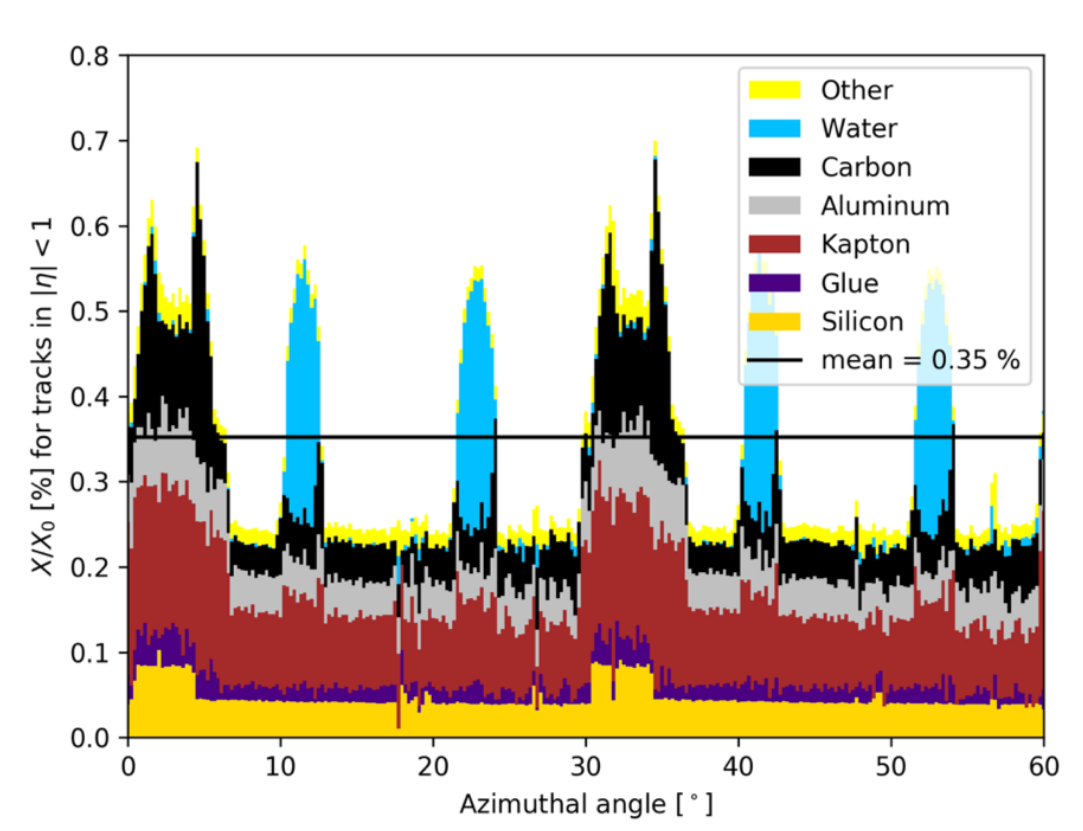}
    \caption{Left: ITS3 layout with 3 half-layer sensors (green) held in place by carbon foam (gray) and surrounded by a cylindrical support structure (beige). The beam pipe (orange) has 16 mm radius to allow layer 0 to be placed at 18 mm from the interaction point. Right: Material budget in current ALICE ITS2 layer 0. By removing material such as kapton and aluminum for circuit boards, water for cooling, and and carbon and glue for mechanical support, wafer-size bent, stitched silicon sensors would result in a material budget of only 0.05\% $X_0$ (the bottom yellow area in this graph). Figures from \cite{loiits3}.}
    \label{fig:bentlayers}
    \label{fig:materialbudget}
\end{figure}

%From ITS2 to ITS3
In 2027, it is foreseen to replace the inner barrel layers of ITS2 with new, truly cylindrical bent sensors, the ITS3. A model is shown on the right in Fig. \ref{fig:er1em1}. The 24120 chips from 200~mm wafers placed at distances down to 22~mm from IP will be replaced with stitched, wafer-scale sensors from 300 mm wafers (see Fig. \ref{fig:stitchedsensor}) bent in half-cylindrical shapes (see Fig. \ref{fig:materialbudget}), with a minimum radius of the innermost layer of 18~mm from the IP. The material budget will be decreased even further down from 0.36\%~$X_0$ per layer to about $0.05\%~X_0$ per layer, as shown on the right in Fig. \ref{fig:materialbudget}. To improve the proximity to IP, the current beam pipe with 
%a central beryllium section of 888~mm and 
an outer radius of 18~mm \cite{alicels2} will be replaced with one of only 16~mm radius and 500 $\mu$m beryllium corresponding to 0.14\%~$X_0$.

An example of a pad wafer from one of the chip submissions, engineering run 1 (ER1), that uses stitching on a 30 cm wafer, is shown in Fig. \ref{fig:stitchedsensor}. In the process of stitching, design blocks are put together during the processing of the silicon. This can make a chip larger than the field of view of the lithographic equipment.% A model for the ITS3 using 3 layers of dummy silicon is shown in Fig. \ref{fig:er1em1}.
%\begin{figure}
%    \centering
%    \includegraphics[width=0.35\textwidth]{engineering_model1.png}
%    \caption{ITS3 engineering model 1 made of 3 layers of 40-50 $\mu$m thick dummy silicon. Figure from \cite{magnusdrd7}.}
%    \label{fig:em1}
%\end{figure}

The ALICE ITS3 upgrade for Run 4 will feature 6 half-layer sensors of 26~cm long in the $z$ direction, two in each layer, that will be thinned to 40-50 $\mu$m. Each sensor will consist of 3-5 wafer-scale stitched MAPS, with one layer 0 sensor spanning 53.3~mm in $r\varphi$. These will be mechanically held in place by carbon foam. The structure that is foreseen is shown on the left in Fig. \ref{fig:bentlayers}. The new beam pipe will allow to place the ITS3 innermost layer 0 at only 18 mm from the interaction point.

\section{ITS3 requirements and performance}
The ITS3 vertex detector has to withstand a fluence resulting from a non-ionizing energy loss of $\Phi_{\mathrm{eq}} = 10^{13}~\mathrm{1~MeV}~\it{n}_{\mathrm{eq}}/\mathrm{cm}^{2}$ and a total ionizing dose of 10 kGy. The particle rates will be up to 2.2 MHz/cm$^2$ in the innermost layer for a Pb--Pb interaction rate of 50 kHz.

\begin{figure}
    \centering
    \includegraphics[width=0.33\textwidth]{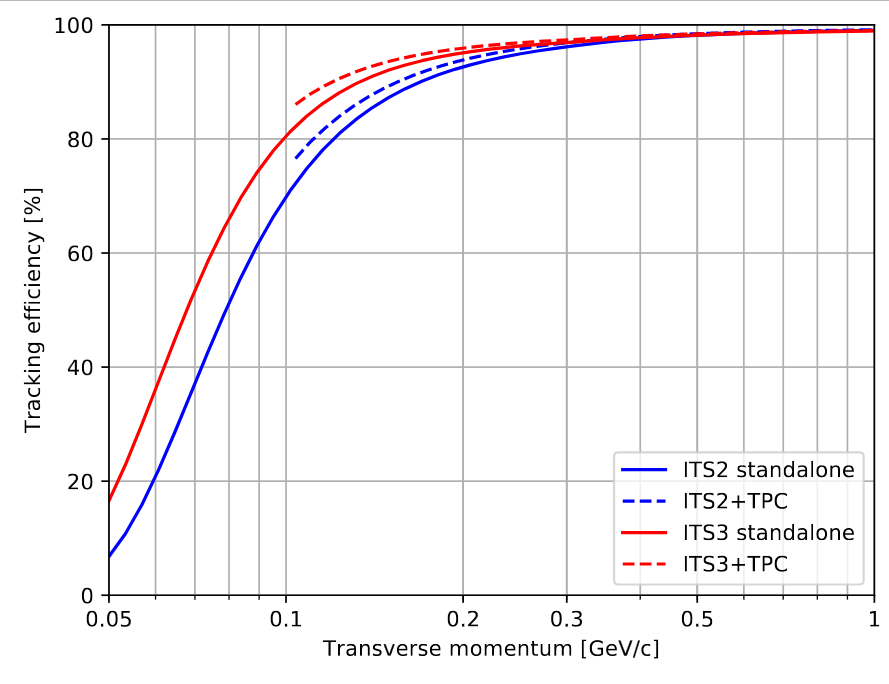}
    \includegraphics[width=0.33\textwidth]{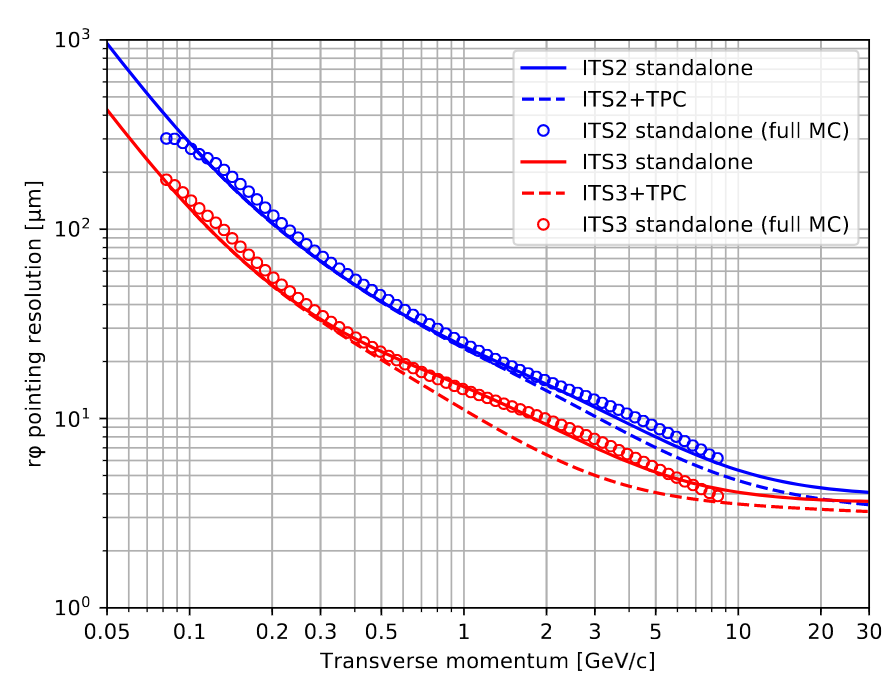}
    \includegraphics[width=0.31\textwidth]{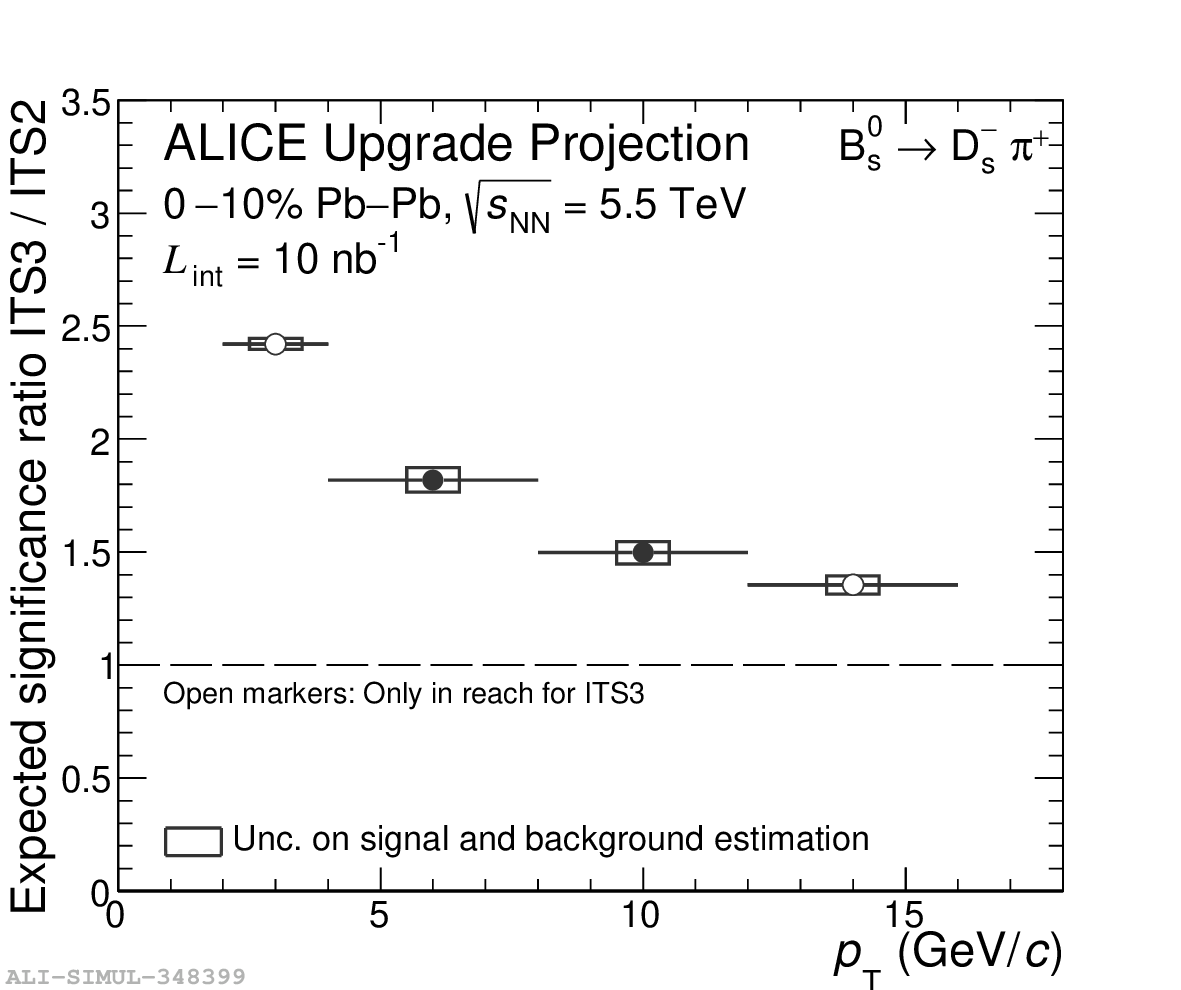}
    \caption{Comparison of the track-reconstruction efficiency (left panel) and pointing resolution in the transverse plane (middle panel) with the current ITS2 and the planned upgrade ITS3 detector using a fast simulation Monte Carlo Tool as well as a full Monte Carlo simulation (circles, middle panel only). Figures from \cite{loiits3}. Right: ratio of the statistical significances expected with the ITS3 and ITS2 for the reconstruction of the $\mathrm{B}^0_{\mathrm{s}} \to \mathrm{D}^-_{\mathrm{s}}\pi^+$ decay in Pb--Pb collisions at $\sqrt{s_{\rm NN}} = 5.5$ TeV. At transverse momenta of around 3 GeV and 14 GeV, the measurement is only possible with the ITS3 upgrade detector.}
    \label{fig:vertexing}
\end{figure}

The ITS3 is expected to improve the tracking efficiency compared to the current vertex detector ITS2, especially at low transverse momentum, as shown on the left in Fig.~\ref{fig:vertexing}. It will also allow a factor of two improvement in the pointing resolution in the $r\phi$ plane over the full range of transverse momenta, as can be seen in the middle of Fig.~\ref{fig:vertexing}. To simulate this, a fast Monte Carlo tool that includes multiple scattering, secondary interactions and detector occupancy was used \cite{loiits3}. This tool ignores the particles’ energy loss in the beampipe and in the detector.

The comparison of the production yields of strange and non-strange hadrons in the heavy-flavor sector is a necessary measurement to understand heavy-flavor hadronization. 
%Strange particles are ideal for studies of hadronization in heavy ion collisions.
In the beauty sector, both ALICE and CMS made first measurements sensitive to the $B_{\rm s}^{0}$-to-non-strange B meson yield ratio, but the uncertainties prevented to conclude about a possible enhancement with respect to pp collisions \cite{alicestrangeness,cmsstrangeness}. 
%The Compact Muon Solenoid (CMS) Experiment at the LHC made a first measurement of $\mathrm{B}^0_{\mathrm{s}}/\mathrm{B}_{\mathrm{not~s}}$ in Pb-Pb collisions as compared to proton-proton collisions. ALICE also measured this quantity \cite{alicestrangeness}. Both ALICE and CMS saw an enhancement in this quantity \cite{cmsstrangeness}, but no significant observation was made.
The ITS3 will allow for a large improvement on this measurement as well as for an extension of the measurement to much lower transverse momenta, as shown on the right in Fig.~\ref{fig:vertexing}. The large improvements in tracking, vertexing, and physics performance expected with ITS3 are all results of the reduced pixel pitch, the very close proximity to the interaction point -- a mere 18 mm -- and the very low material budget.
\section{Reduced material budget}
The very low material budget, which contributes to the excellent expected performance of the ITS3 vertex detector, results from removing all the ``unnecessary'' material in the current ALICE inner tracking system. Circuit boards with kapton and aluminum are, for example, not required if power distribution and data transmission are integrated into the silicon. This is achieved with long, stitched, wafer-scale sensors. These same large-area sensors that are bent around the beam pipe also need less mechanical support, reducing the amount of carbon and glue needed. Finally, if power consumption of these sensors can be kept below 20 mW/cm$^2$, water cooling can be replaced by air cooling, further reducing the amount of material budget. Overall, as can be deduced from Fig. \ref{fig:materialbudget}, the material budget per layer of 0.35\% $X_0$ for the current ITS2 can be reduced to 0.05\% for the ITS3.

\begin{figure}
    \centering
    \includegraphics[width=0.18\textwidth]{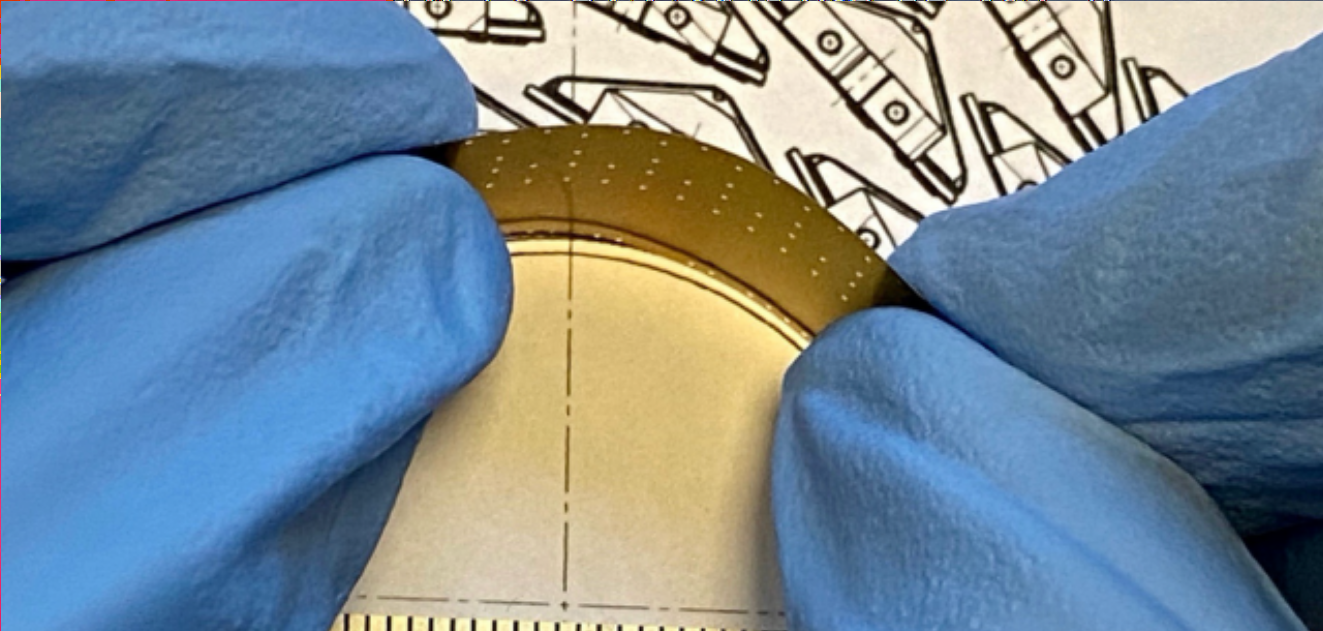}
    \includegraphics[width=0.79\textwidth, trim=0 0 0 378, clip]{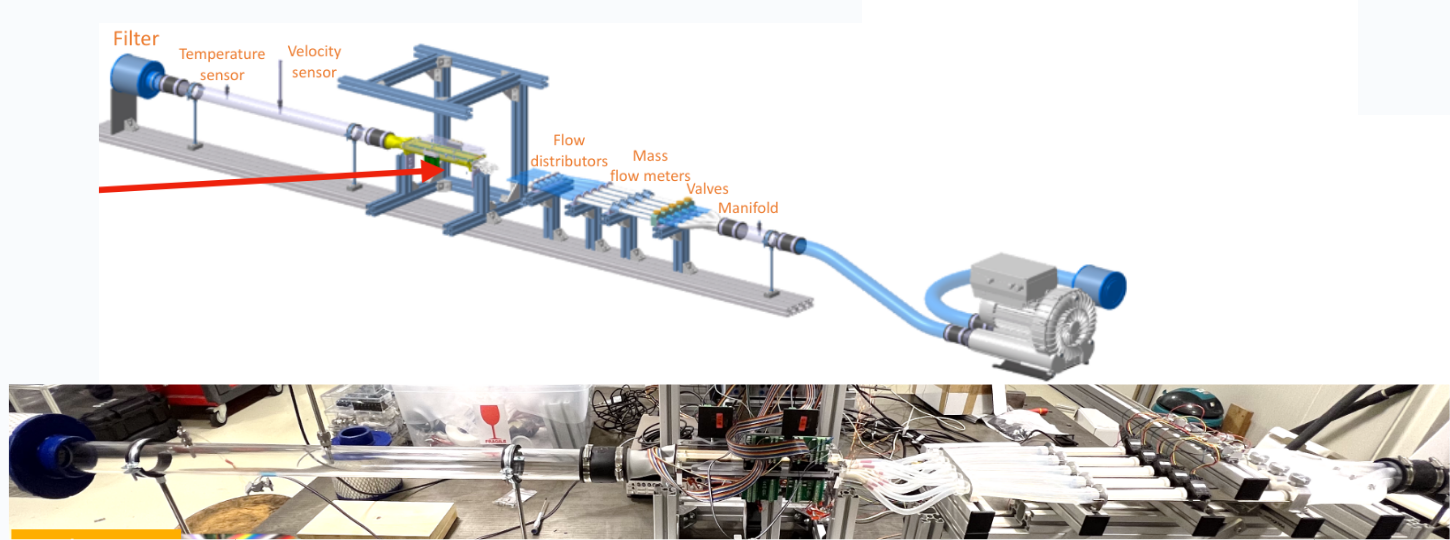}
    \caption{
        Left: an ALPIDE monolithic active pixel sensor bent by hand.
    Right: air cooling is being extensively studied in a commissioned setup at CERN.}
    \label{fig:bent_aircooling}
\end{figure}

\begin{figure}
    \centering
    \includegraphics[width=0.274\textwidth, trim=500 500 700 500, clip]{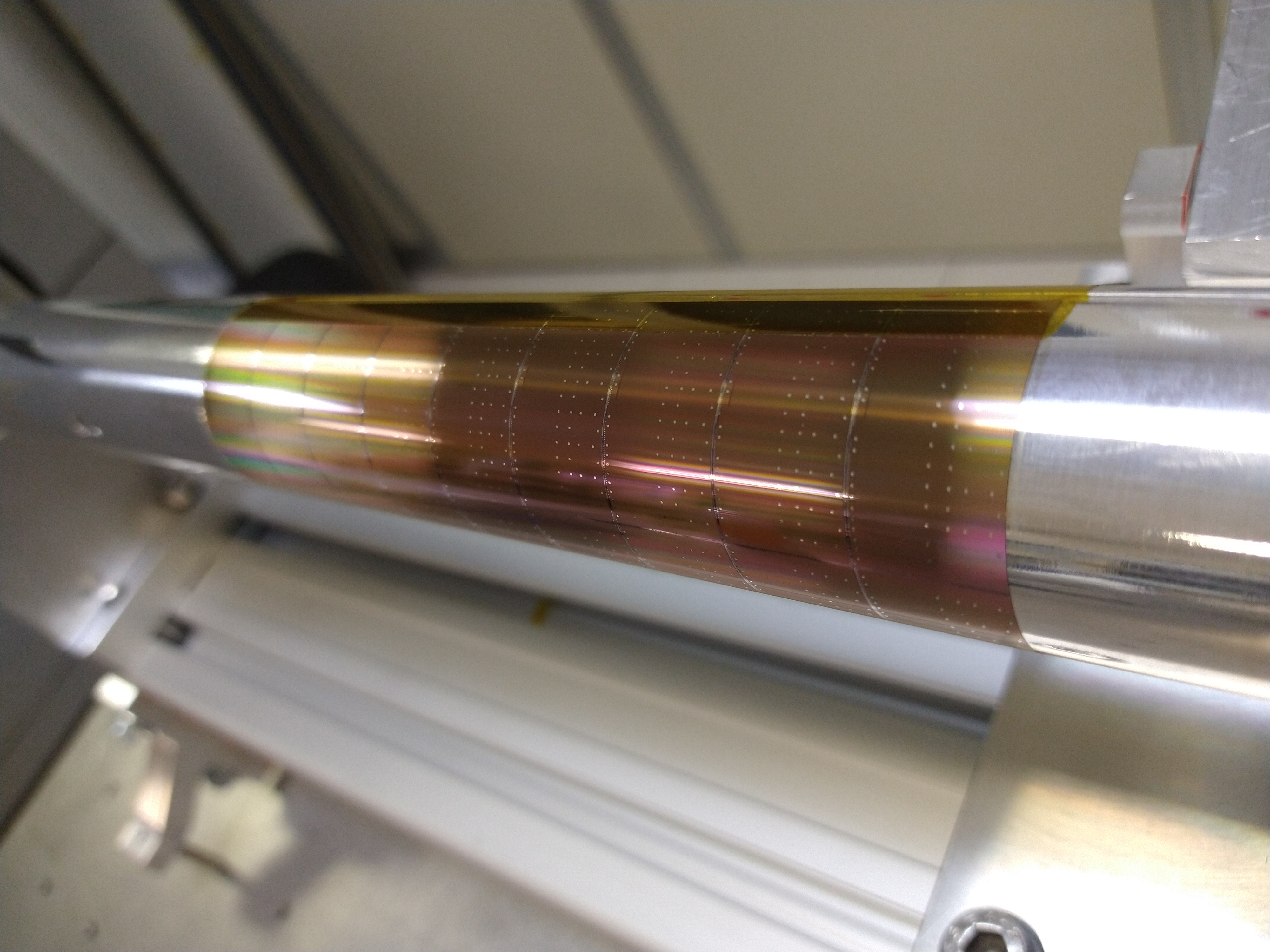}
    \includegraphics[width=0.274\textwidth, trim=500 500 500 500, clip]{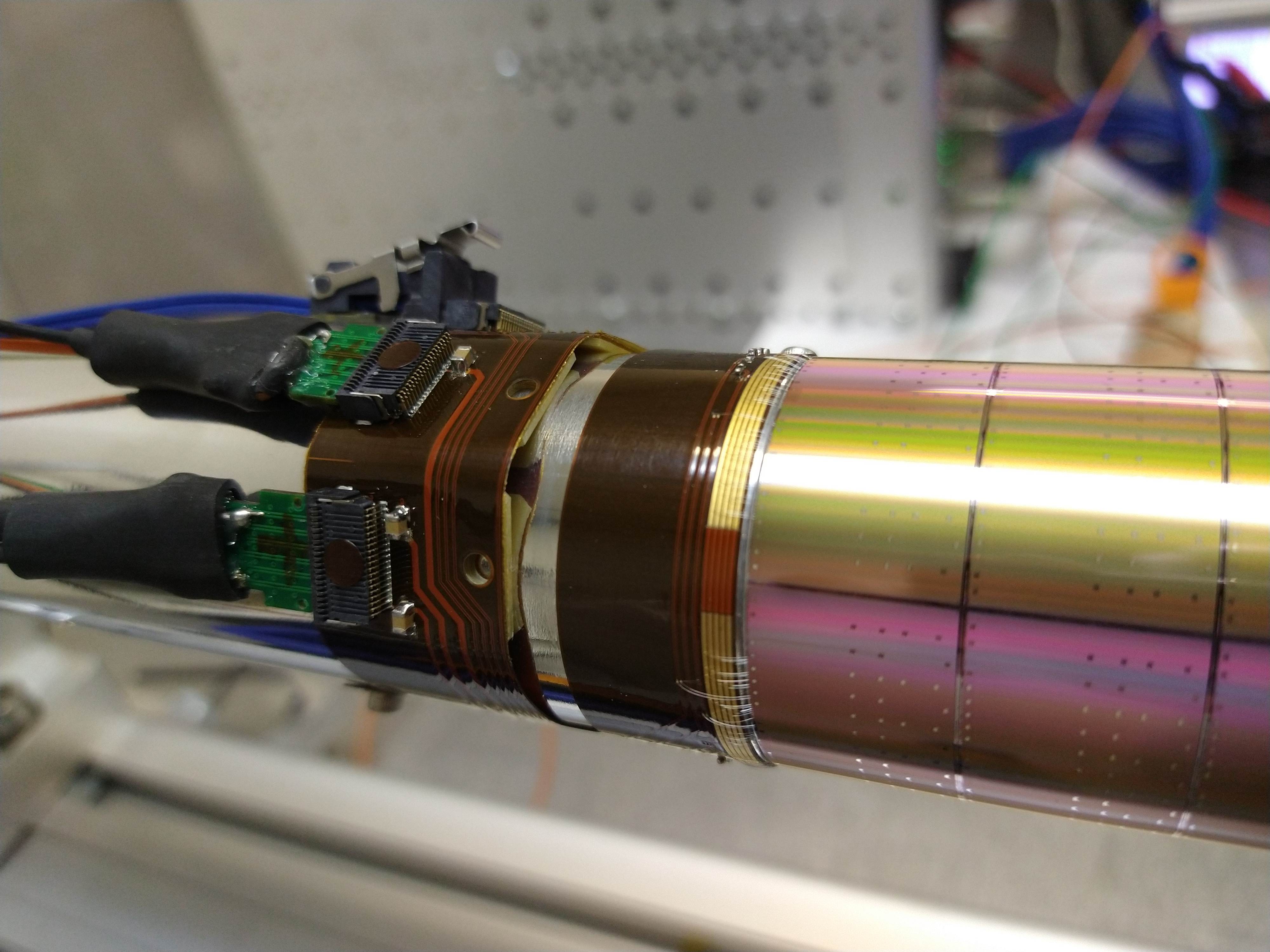}
    \includegraphics[width=0.16\textwidth]{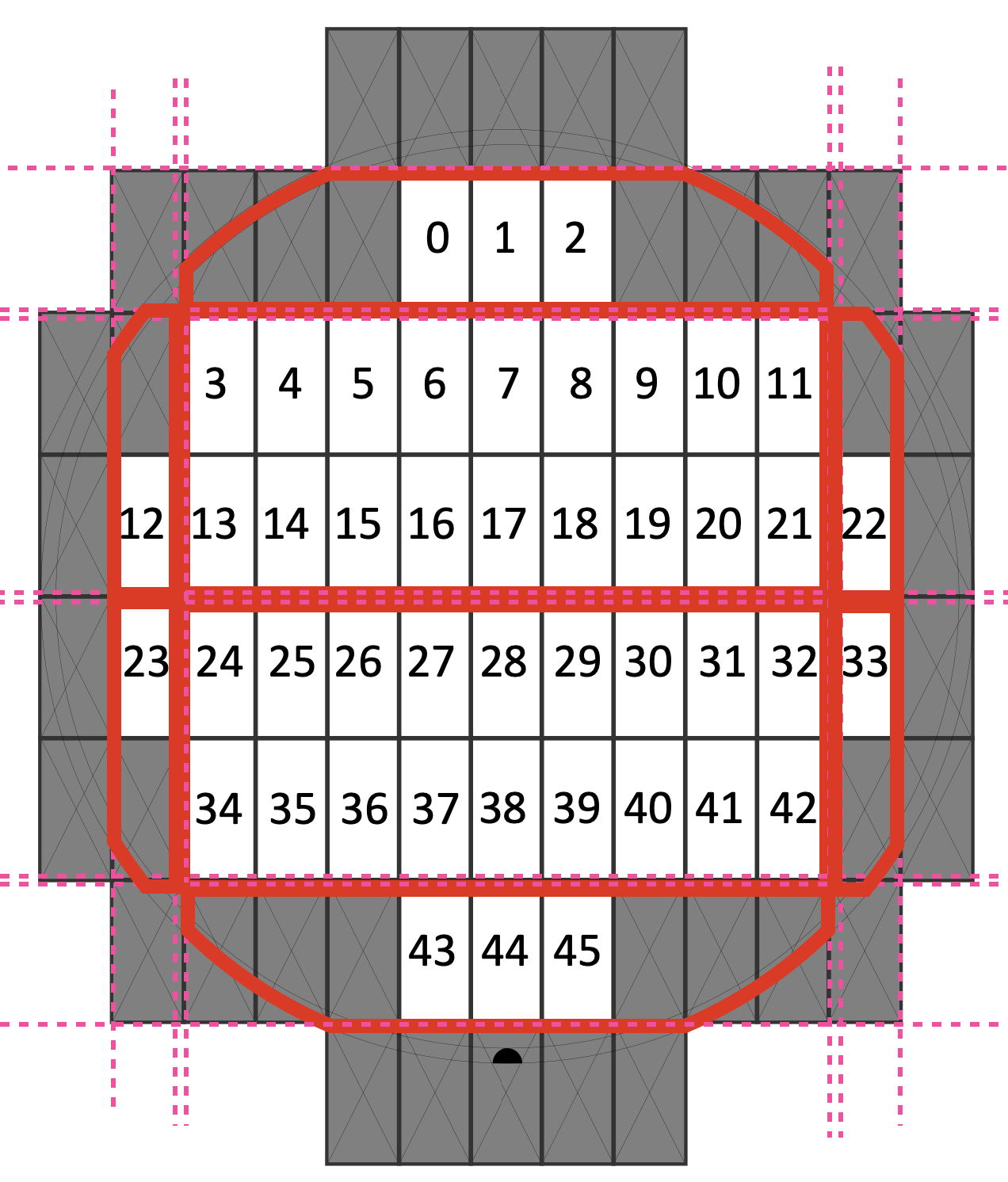}
    \includegraphics[width=0.23\textwidth, trim=980 1380 1380 180, clip]{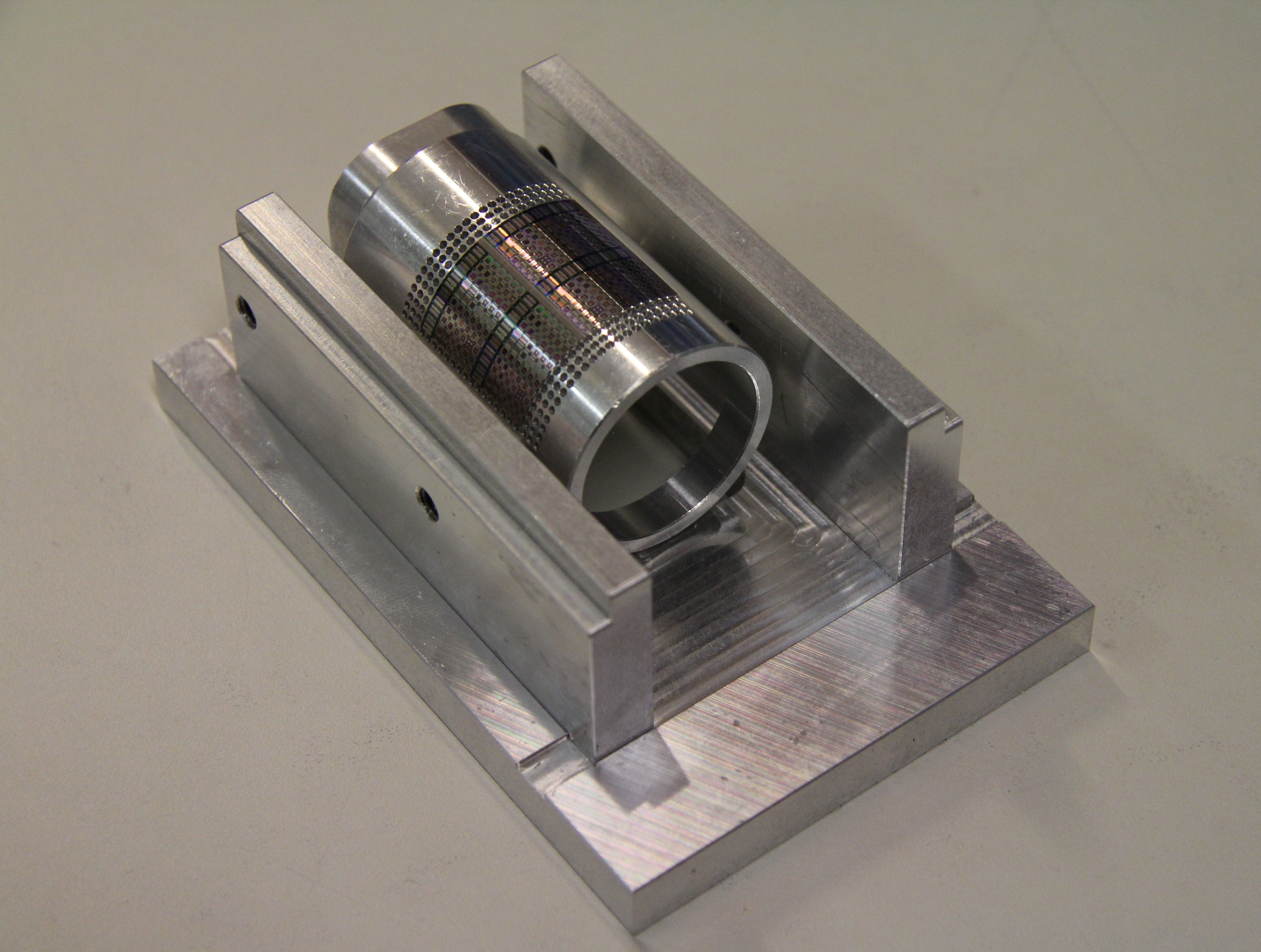}
    \caption{Bent ``SuperALPIDEs'' (left), or multiple ALPIDEs that were produced for ITS2 from a single wafer (schematic, middle right). Bending tests have proven to be successful with this chip as well as with regular ALPIDEs bent and studied in beam test facilities. Right: a 65 nm prototype for ITS3 has also been successfully bent and tested.}
    \label{fig:superalpidebent}
\end{figure}

To test the concept of air cooling, a setup with a wind tunnel and a laser measurement system has been commissioned, as shown in Fig. \ref{fig:bent_aircooling}. Mechanical and stability tests are ongoing. The bending of the silicon sensors is studied extensively, with first tests of 40 $\mu$m thick ``superALPIDEs'', or multiple ALPIDEs from a wafer that were not cut, having proven to show successful bending to 18 mm, as shown in Fig. \ref{fig:superalpidebent}. Subsequent beam tests have proven that the ALPIDE in an ITS3 mock-up called the $\mu$ITS3 with sensors bent to the ITS3 radii of 18, 24, and 30 mm show an efficiency and resolution consistent with flat ALPIDEs \cite{bentalpide}. The spatial resolution was also uniform across different radii.

\section{Sensor R\&D}
\begin{figure}
    \centering
    \includegraphics[width=0.847\textwidth]{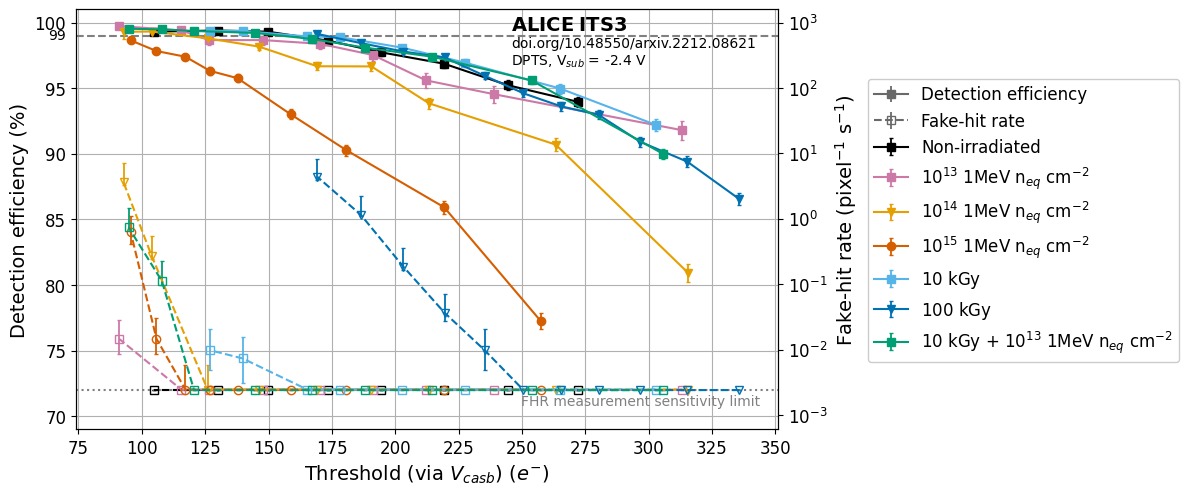}
    \caption{
        Detection efficiency and fake-hit rate for non-irradiated sensors and for sensors irradiated with different fluences and dose. This 65 nm digital pixel test structure was shown to be 100\% efficient at room temperature after the ITS3 expected radiation load and still operable at 99\% efficiency after a fluence of 100 times this expected load. Figure from \cite{dpts}.}
    \label{fig:dpts}
\end{figure}

Large-scale wafers of 30~cm and the process of stitching are available in a Tower Partners Semiconductor Co (TPSCo) 65 nm technology. As the ALPIDEs were produced in a 180 nm CMOS imaging process provided by TowerJazz \cite{alpide}, active sensor research and development is ongoing for the new 65 nm technology. There are several submissions planned, and a prototype of a final wafer-scale chip is expected at the end of 2024. Characterization of many small prototypes from a first prototype run on a multi-layer reticle proved operation at room temperature of $20~^{\circ}$C of a digital pixel test structure to be 100\% efficient after being irradiated with the ITS3 expected fluence and dose, and operable at 99\% efficiency after a fluence of 100 times that of $\Phi_{\mathrm{eq}} = 10^{15} ~\mathrm{1~MeV}~\it{n}_{\mathrm{eq}}/\mathrm{cm}^{2}$
\cite{dpts}, as shown in Fig. \ref{fig:dpts}. This digital pixel test structure has 32 $\times$ 32 pixels of 15 $\mu$m pitch whose position is time-encoded in an asynchronous digital readout. Charge loss occurs in the corners of a pixel far from the collection electrode, as expected from the very little to no charge sharing. The spatial track resolution was determined to be 2.4 $\mu$m. The first structures bent to a radius of 18 mm, as shown on the right in Fig. \ref{fig:superalpidebent}, were successfully tested in the laboratory with an $^{55}$Fe source.

The first stitched sensor prototypes are now being investigated. A layout is shown in on the left in Fig. \ref{fig:er1em1}. There are two different structures, the monolithic stitched sensor (MOSS) of 14 $\times$ 259 mm$^{2}$ with $6.72\times 10^6$ pixels, and the monolithic stitched sensor with timing (MOST) of 2.5 $\times$ 259 mm$^2$ with $0.9 \times 10^6$ pixels. The full structure for the ITS3 will be 2.5 times as large. Pixels of both 18 and 22.5 $\mu$m pitch are available. The structures will be tested for uniformity and yield.

\section{Summary and outlook}
The ALICE collaboration plans the installation of new inner layers (ITS3) in 2027 for the ALICE inner tracking system for LHC Run 4. The aim is to use truly cylindrical wafer-scale monolithic active pixel sensors. Silicon flexibility and bending have been proven with routine tests and a full mock-up of the ITS3 was shown to be efficient when bent to the ITS3 target radii.

For access to wafer-scale stitched sensors, a new 65 nm CMOS imaging technology is used. The first prototypes reach 100\% detection efficiency at room temperature at the ALICE ITS3 expected fluence of  $\Phi_{\mathrm{eq}} = 10^{13}
~\mathrm{1~MeV}~\it{n}_{\mathrm{eq}}/\mathrm{cm}^{2}$
, and the sensor is still operable at room temperature after  $\Phi_{\mathrm{eq}} = 10^{15}
~\mathrm{1~MeV}~\it{n}_{\mathrm{eq}}/\mathrm{cm}^{2}$. The first stitched sensors are being tested now. The ITS3 R\&D will pave the way to thin, low-power sensors that could be used in future experiments like ALICE3 \cite{alice3} and will enable a wealth of new precision measurements.

\end{document}